\documentclass[10pt]{iopart}
\usepackage{epsfig,cite}

\begin{document}
\input macros.sty
\begin{flushright}
\vspace{-3.1cm}
       {\normalsize\tt CERN-TH/2000-245}
\end{flushright}
       \vspace{0.7cm}
\title{Strangeness in Lattice QCD}

\author{Hartmut Wittig\footnote{On leave of absence from: Theoretical
    Physics, University of Oxford, 1~Keble Road, Oxford
    OX1~3NP, UK\\
    Invited talk presented at V International Conference on
    Strangeness in Quark Matter, ``Strangeness 2000'', Berkeley, CA,
    20--25 July 2000}}

\address{CERN, Theory Division, CH-1211 Geneva 23, Switzerland}

\begin{abstract}
  The status of lattice calculations in the light hadron sector is
  reviewed. Special emphasis is given to recent lattice determinations
  of the mass of the strange quark. The impact of non-perturbative
  renormalization and control over lattice artefacts on the attainable
  precision is discussed in detail. Furthermore the influence of
  dynamical quark effects is assessed.
\end{abstract}


\section{Introduction}

Quantum Chromodynamics (QCD) is widely accepted as the theory
describing the strong interaction. It is formulated in terms of quarks
and gluons and contains relatively few free parameters, namely the
gauge coupling $g$ and the masses $\mup,\,\md,\,\mstr,\ldots$ of the
various quarks. These parameters must be fixed by using experimental
input: a typical example is the determination of the sum of quark
masses $(\mup+\mstr)$, using the experimental value of the kaon mass,
$m_{\rm K^+}$:
\be
   m_{\rm K^+} \mapsto (\mup+\mstr).
\ee
Clearly, in order to connect the two sides of this relation one has to
``solve QCD'' for the case at hand.

The QCD coupling constant $g$, which describes the coupling strength,
depends on the momentum transfer~$q$ between quarks and gluons. The
well-known property of ``asymptotic freedom'' implies that~$g$
decreases for large~$q$. By contrast, the coupling becomes large when
momentum transfers of the order of typical hadronic scales are
considered, say, $q\approx1\,\gev/c$. As a consequence, perturbation
theory in $g^2$ becomes an inadequate tool to study QCD at low
energies. Hence, a non-perturbative treatment is required in order to
understand hadronic properties on a quantitative level. The
formulation of QCD on a (Euclidean) space-time lattice~\cite{Wilson74}
provides such a non-perturbative method to compute the relations
between experimentally accessible quantities and the parameters of QCD
through numerical simulations.

Here I restrict myself to QCD at zero temperature, focusing on the
spectrum of light hadrons and the determination of the strange quark
mass from first principles. A more general overview of the status of
lattice calculations can be obtained from the proceedings of the
annual Conference on Lattice Field Theory~\cite{proc_lat99} and other
review talks~\cite{kenway00,aoki99,eps99,sharpe_ichep98}.

In the next section I shall outline the basic concepts of the lattice
formulation. Section~\ref{sec_hadrons} contains results from recent
benchmark calculations of the light hadron spectrum. In
section~\ref{sec_quarks} recent determinations of the strange quark
mass are described. Finally, section~\ref{sec_concl} contains some
concluding remarks.

\section{Basic concepts \label{sec_basic}}

The lattice formulation of QCD replaces the familiar continuous
Minkowski space-time by a discretized, Euclidean version with finite
volume $L^3\cdot{T}$. Points in space-time are separated by a finite
distance~$a$, the lattice spacing. The inverse lattice spacing,
$a^{-1}$, acts as an UV cutoff, which regularizes the infinities that
are typically encountered in Quantum Field Theory. The quark and
antiquark fields $\psi(x),\,\psibar(x)$ are associated with the sites
of the lattice, whereas the gauge field is represented by the
so-called link variable $U_\mu(x)$, which connects neighbouring
lattice sites, and is an element of the gauge group SU(3). After
choosing a suitable, gauge-invariant discretization of the QCD action
\be
   S[U,\psibar,\psi] = \Sg[U]+\Sf[U,\psibar,\psi],
\ee
where $\Sg$ is the lattice action of pure Yang--Mills theory and $\Sf$
denotes the lattice fermion action, one may define the expectation
value of an observable $\Omega$ as
\be
  \langle\Omega\rangle = \frac{1}{Z} \int\,D[U]D[\psibar]D[\psi]
  \,\Omega\,\exp\left(-\Sg-\Sf\right).
\ee
Here the normalization of the functional integral~$Z$ is determined by
requiring $\langle\unt\rangle=1$. After performing the integration
over fermionic fields the expression for the expectation value becomes
\be
  \langle\Omega\rangle = \frac{1}{Z} \int\prod_{x,\mu}dU_\mu(x)
  \prod_{\rm f}\det\left(D+\mf\right)\,\Omega\,\exp(-\Sg),
  \label{eq_expval}
\ee
where $D$ is the lattice Dirac operator and $\mf$ is the mass of quark
flavour~$f$. Thus, the discretization procedure has given a meaning to
the functional integration over gauge fields, which reduces to an
ordinary multiple-dimensional integral over group elements. The
various steps leading to \eq{eq_expval} can be treated in a
gauge-invariant manner, and hence the lattice formulation represents a
regularization procedure that preserves the gauge invariance of QCD.
Furthermore, the definition of physical observables does not rely on
perturbation theory. Therefore \eq{eq_expval} forms the basis for a
non-perturbative, {\it stochastic} evaluation of expectation values of
physical observables using numerical simulations.

Despite an enormous increase in computer power, there remain several
major difficulties that make {\it realistic} simulations of QCD a hard
task. Perhaps the biggest challenge is the inclusion of dynamical
quark effects: the evaluation of the fermionic determinant
in~\eq{eq_expval} in numerical simulations is still very costly, even
on today's massively parallel computers. In early simulations of
lattice QCD the determinant was therefore set equal to~1, a choice
whose physical interpretation corresponds to neglecting quark loops in
the evaluation of $\langle\Omega\rangle$. Although this represents a
rather drastic assumption about the influence of quark-induced quantum
effects, the quenched approximation works surprisingly well, as we
shall see later. However, it is clearly desirable to develop
simulation algorithms that allow for a more efficient evaluation of
the fermionic determinant. With present algorithms, the cost of
simulating ``full'' QCD is roughly a thousand times higher than that
of the corresponding quenched simulation.

An indirect consequence of using the quenched approximation is the
so-called {\em scale ambiguity}. That is, the calibration of the
lattice spacing in physical units, $a^{-1}\,[\mev]$, depends on the
quantity $Q$, which is used to set the scale
\be
   a^{-1}\,[\mev] = \frac{Q\,[\mev]}{(aQ)},\quad Q=f_\pi,m_N,
   m_\rho,\ldots
\ee
This ambiguity arises because different quantities~$Q$ are affected by
quark loops in different ways. Clearly, dimensionful quantities such
as particle masses are directly affected by the scale ambiguity.

Another problem one has to address are lattice artefacts (cutoff
effects). Let $\Omega$ denote the mass of a hadron in units of, say,
the nucleon mass. Then the expectation values on the lattice and in
the continuum differ by corrections of order $a^p$:
\be
  \langle\Omega\rangle^{\rm lat} = \langle\Omega\rangle^{\rm cont}
  +O(a^p),
\label{eq_o_cont_lat}
\ee
where the value of integer~$p$ in the correction term depends on the
chosen discretization of the QCD action. Values of~$a$ that can
currently be simulated lie in the range $a\approx0.2$--$0.05\,\fm$.
The size of the correction term can in some cases be as large as
$20$\%, depending on the quantity and the chosen discretization. An
extrapolation to the continuum limit, $a\to0$, is then required to
obtain the desired result. Surely this extrapolation is much better
controlled if the discretization avoids small values of~$p$.

Finally there are restrictions on the quark masses $\mf$ that can be
simulated. In general the following inequalities should be satisfied
in any simulation:
\be
  a\ll\xi\ll L,
\ee
where~$L$ is the spatial extent of the lattice volume. The
quantity~$\xi$ denotes the correlation length of a typical hadronic
state and serves as a measure of the quark mass. The inequality on the
right places restrictions on the light quark masses that can be
simulated: if those are too light one may suffer from finite-size
effects, since~$\xi$ becomes large. Typical spatial extensions of
$L\approx1.5$--$3\,\fm$ imply that the physical pion mass cannot be
reached. The left inequality restricts the masses of heavy quarks.
Since $a^{-1}\approx2$--$4\,\gev$, it is clear that relativistic
$b$-quarks cannot be simulated. One therefore relies on extrapolations
in $\mf$ to connect to the physical $u,\,d$ and~$b$ quarks.

\section{Hadron spectroscopy \label{sec_hadrons}}

In many ways the calculation of the mass spectrum of light hadrons is
a benchmark of lattice QCD. A comparison of the experimentally
observed spectrum with the results from an ``ideal'' lattice
simulation of QCD, in which dynamical quark effects, lattice artefacts
and quark mass dependences are all sufficiently controlled, would
represent a stringent test of QCD as the underlying theory of the
strong interaction. Alternatively, such a comparison enables us to
assess the inherent systematics of current simulations, notably the
effects of dynamical quarks. In particular one may investigate the
quality of the quenched approximation.

\subsection{Quenched light hadron spectrum}

\begin{figure}[tb]
\vspace{1.cm}
\hspace{3.cm}
\ewxy{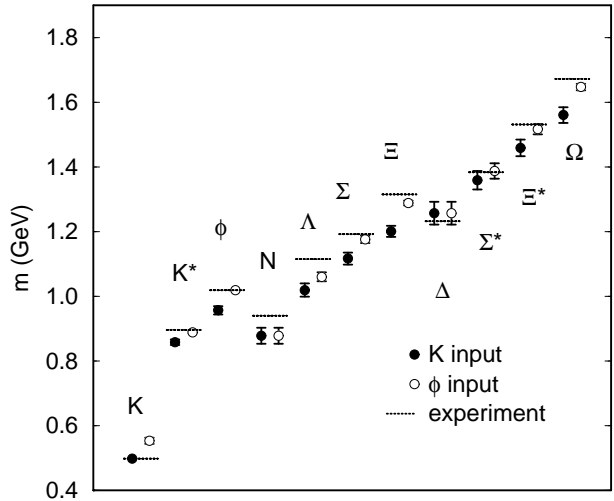}{100mm}
\vspace{-4.5cm}
\caption{The quenched hadron spectrum from
ref.~\protect\cite{CP-PACS_quen} compared with experiment (dashed
lines).}
\label{fig_CPPACS_quen}
\end{figure}

Recently the CP-PACS Collaboration presented a precision calculation
of the light hadron spectrum in quenched QCD~\cite{CP-PACS_quen},
which superseded an earlier study by GF11~\cite{GF11_quen}. The
findings of CP-PACS are summarized in the plot shown in
Fig.~\ref{fig_CPPACS_quen}. The main features of the spectrum are well
reproduced by the quenched lattice data. Bearing in mind that only two
input quantities have been used, namely the mass of the $\rho$-meson
to set the scale and the mass of a strange meson (either $\mK$ or
$m_\phi$) to fix the mass of the strange quark, the fact that the
masses of so many hadrons are predicted quite accurately represents a
major achievement in the understanding of QCD. Nevertheless, one finds
small but significant deviations from the experimentally observed
spectrum. For instance, the ratio of the nucleon and $\rho$ masses is
calculated as
\be
  m_{\rm N}/m_\rho = 1.143\pm0.033,
\ee
which is 6.7\% (2.5$\sigma$) below the experimental value of~1.218.
Similarly, vector-pseudoscalar mass splittings such as $m_{\rm
  K^*}-\mK$ are too small by $10$--$16$\% (4--6$\sigma$), depending on
whether $\mK$ or $m_\phi$ is used to fix the strange quark mass. This
implies that, for the first time, a {\it significant\/} deviation
between the quenched QCD spectrum and nature is detected. Thus the
conclusion is that quenched QCD describes the light hadron spectrum at
the level of~10\%. However, it also shows that the quenched
approximation works surprisingly well, since the discrepancy is fairly
mild. This has important consequences for lattice predictions of some
phenomenologically interesting quantities, for which one has to rely
on the quenched approximation for some time.

Although the CP-PACS results represent a real benchmark in terms of
statistics, parameter values and lattice volumes, further
corroboration of these findings is required. Recent calculations
employing different discretizations have largely confirmed the
findings of~\cite{CP-PACS_quen}: the MILC
Collaboration~\cite{MILC_quen} has used staggered fermions~\cite{KS75}
and finds a value for $m_{\rm N}/m_\rho$ in the continuum limit of
\be
 m_{\rm N}/m_\rho = 1.254\pm0.018\,{\rm(stat)}\pm0.028\,{\rm (syst)}.
\ee
This is in broad agreement with experiment, but the difference to the
CP-PACS result amounts to only 2$\sigma$. A recent calculation by
UKQCD using $O(a)$ improved Wilson fermions~\cite{UKQCD_quen} finds
$m_{\rm N}/m_\rho = 1.26^{+\phantom{1}8}_{-14}$. Whereas the error is
too large to detect a significant deviation, UKQCD's results for the
spectrum also indicate that the quenched light hadron spectrum agrees
with experiment at the level of~10\%.

\subsection{Beyond the quenched approximation}

An obvious question is whether sea quark effects can account for the
observed deviation of the quenched light hadron spectrum from
experiment. Recently several collaborations have studied the light
hadron spectrum using $\Nf=2$ flavours of dynamical quarks, which are
identified with the physical up and down
quarks~\cite{SESAM_nf2_98,UKQCD_c176,joyce_lat99,kaneko_lat99,colum_00}.
However, the masses of the dynamical quarks that can be simulated are
quite large. A measure for the difference between the light quark
masses found in nature and those used in dynamical simulations is
provided by the pseudoscalar-to-vector mass ratio, $\mps/\mv$. Current
simulations are typically performed for $\mps/\mv=0.6$--0.8, whereas
the physical value is $\mps/\mv=m_\pi/m_\rho=0.169$. Therefore one
relies on extrapolations in the sea quark mass to make contact with
the physical case. In addition, it has not been possible to simulate a
third dynamical quark with $\mf\approx\mstr$ efficiently using known
algorithms. Therefore, in most dynamical calculations an unphysical
number of sea quarks has been used.

Despite these shortcomings one can perform systematic studies of the
influence of dynamical quarks on the light hadron spectrum.
CP-PACS~\cite{kaneko_lat99} have studied the continuum limit of hadron
masses computed for $\Nf=2$ and compared it with the results of the
quenched light hadron spectrum discussed before. They find that the
discrepancy with experiment is reduced from 6.5\% to 1.4\% for
$m_\phi$ computed with $\Nf=2$. For $m_{\rm K^*}$ the gap decreases
from 4.4\% to 1.0\%. This demonstrates clearly that the
experimentally observed spectrum is reproduced more closely when sea
quarks are ``switched on''. Furthermore, results for the
vector-pseudoscalar mass splitting reported by UKQCD~\cite{UKQCD_c176}
also show that lattice data for this quantity approach the
experimental value as the sea quark mass is decreased.

Of course, the small remaining differences between QCD with $\Nf=2$
and experiment have to be explained. It is reasonable to assume that
the unphysical value of~$\Nf$ will have some influence. Further
studies are also required to decide whether there is yet sufficient
control over the extrapolations in the quark masses and those to the
continuum limit.

\section{The mass of the strange quark \label{sec_quarks}}

Quark masses are important input parameters in many theoretical
applications but, despite a great deal of activity, relatively little
is known about their absolute values~\cite{PDG2000}. Here I shall
describe calculations of current quark masses defined through the PCAC
relation. For charged kaons it can be written as
\be
  \fK\mK^2 = (\mbaru+\mbars) \langle0|\overline{u}\gamma_5{s}|
  {\rm K}\rangle,
\label{eq_PCAC}
\ee
where the bars above the quark masses indicate that the ``running''
masses are considered, which depend on the energy scale. In order to
determine the sum of up and strange quark masses using the
experimental result for $\fK\mK^2$ one simply has to compute the
matrix element $\langle0|\overline{u}\gamma_5{s}|{\rm K}\rangle$ in a
lattice simulation. However, in order to represent meaningful
theoretical input, quark masses have to be renormalized, and hence
their values depend on the adopted renormalization procedure. By
convention quark masses are quoted in the $\MSbar$ scheme of
dimensional regularization at a reference scale,
typically~$\mu=2\,\gev$.  This implies that the renormalized matrix
element in the $\MSbar$ scheme has to be determined; it is related
to its lattice counterpart by
\be
   \langle0|\overline{u}\gamma_5{s}(\bar{\mu})|{\rm K}\rangle_\MSbar
   = \zp^\MSbar(g_0,a\bar{\mu})
   \langle0|\overline{u}\gamma_5{s}|{\rm K}\rangle_{\rm lat}.
\ee
Here $g_0$ is the bare coupling, and~$\bar{\mu}$ is the subtraction
point in the $\MSbar$ scheme. The renormalization factor~$\zp^\MSbar$
is known to one-loop order in lattice perturbation theory. However, it
is well known that lattice perturbation theory converges slowly, and
in order to remove all doubts about the reliability of the matching
procedure it is evident that a non-perturbative determination of the
renormalization factor is required.

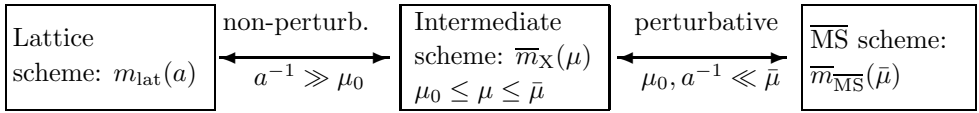
\begin{figure*}[t]
\begin{center}
\unitlength 0.92cm
\begin{picture}(14,2.0)
\put(0.0,0.0){
   \framebox(3.0,1.5){
     \parbox{26mm}{\sloppy
       Lattice\\[0.3ex] scheme: $m_{\rm lat}(a)$
     }
  }
}
\put(5.7,0.0){
   \framebox(3.0,1.5){
     \parbox{24mm}{\sloppy
       Intermediate\\[0.2ex] scheme: $\mbar_{\rm X}(\mu)$
   \\[0.2ex] $\mu_0\leq\mu\leq\bar{\mu}$

     }
  }
}
\put(11.5,0.0){
   \framebox(2.5,1.5){
     \parbox{21mm}{\sloppy
       $\MSbar$ scheme:\\[0.3ex] $\mbar_{\MSbar}(\bar{\mu})$
     }
  }
}
\thicklines
\put(3.2,0.75){\vector(1,0){2.45}}
\put(5.6,0.75){\vector(-1,0){2.4}}
\put(3.25,1.15){\mbox{non-perturb.}}
\put(3.7,0.35){\mbox{${a}^{-1}\gg\mu_0$}}
\put(8.95,0.75){\vector(1,0){2.45}}
\put(11.35,0.75){\vector(-1,0){2.4}}
\put(9.2,1.15){\mbox{perturbative}}
\put(9.3,0.35){\mbox{$\mu_0,a^{-1}\ll\bar{\mu}$}}
\end{picture}
\end{center}
\vspace{-0.3cm}
\caption{Schematic relation between quark masses in lattice
regularization and the $\MSbar$ scheme through an intermediate
renormalization scheme~$\rm X$.}
\label{fig_intscheme}
\end{figure*}

However, a non-perturbative normalization condition relating the
matrix elements in lattice regularization and the $\MSbar$ scheme
cannot be formulated, since the latter is only defined to any given
order of perturbation theory. This technical difficulty can be
overcome by introducing an intermediate scheme~$\rm X$ and considering
a two-step matching procedure, as shown schematically in
Fig.~\ref{fig_intscheme}. The first part is the matching of the bare
current quark mass $m_{\lat}(a)$ to the running mass in the
intermediate scheme $\overline{m}_{\rm X}(\mu)$. This amounts to
computing~$\zp^{\rm X}(g_0,a\mu_0)$ between the lattice scheme and
scheme~X at a fixed scale~$\mu_0$. The second part is the
determination of the scale dependence of the running
mass~$\overline{m}_{\rm X}(\mu)$ from~$\mu_0$ up to very high
energies, where the perturbative relation
between~$\overline{m}_{\rm{X}}$ and~$\mbar_\MSbar$ is expected to be
reliable. Through this two-step process the use of lattice
perturbation theory is completely avoided.

Two examples for such intermediate schemes have been proposed. The
so-called ``regularization-independent'' (RI) scheme is described
in~\cite{MPSTV94}. Another intermediate scheme is defined using the
Schr\"odinger Functional (SF) of
QCD~\cite{LNWW92,sint_SF,martin_LesHouches}. This scheme allows one to
compute the scale dependence non-perturbatively over several orders of
magnitude, using a recursive finite-size scaling technique. Once the
scale dependence of the running mass in the SF scheme, $\mbar_{\rm
  SF}$, is known up to $\mu\approx100\,\gev$ one can continue the
scale evolution to infinite energy using the perturbative
renormalization group functions and thereby extract the
renormalization group invariant (RGI) quark mass~$M$. At this point
the matching to the $\MSbar$ scheme is trivial, since~$M$ is
scheme-independent, i.e. it coincides in the SF and $\MSbar$ schemes.

\begin{figure}[tb]
\vspace{-1.5cm}
\hspace{3.cm}
\ewxy{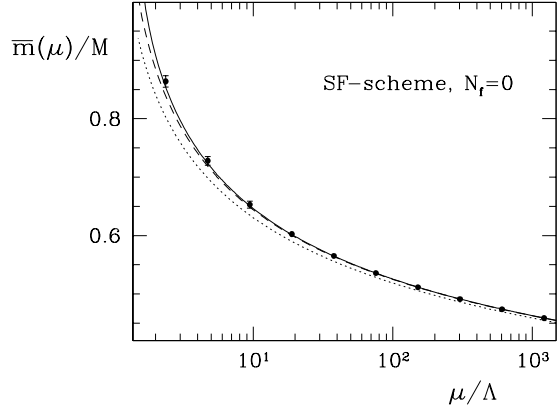}{98mm}
\vspace{-3.2cm}
\caption{Non-perturbative scale evolution of $\mbar_{\rm SF}/M$
computed in lattice simulations of the SF (solid circles). The lines
correspond to the scale evolution computing using various orders of
perturbation theory\label{fig_mbarSF}.}
\end{figure}

Figure~\ref{fig_mbarSF} shows the non-perturbatively determined scale
dependence of $\mbar_{\rm SF}/M$ computed in quenched QCD~\cite{mbar1}
and compares it to the perturbative scale evolution.
The left-most data point in Fig.~\ref{fig_mbarSF} corresponds to a
scale $\mu_0=275\,\mev$. It is here that the matching between lattice
regularization and the $\MSbar$ scheme via the SF is completed,
through a non-perturbative calculation of~$\zp^{\rm SF}(g_0,a\mu_0)$
at fixed $\mu_0=275\,\mev$. All dependence on the intermediate SF
scheme and the scale~$\mu_0$ drops out in the total renormalization
factor
\be
   \frac{\mbar_{\rm SF}(\mu_0)}{M}\times\zp^{\rm SF}(g_0,a\mu_0),
\ee
where $\mbar_{\rm SF}(\mu_0)/M=0.864\pm0.011$, as can be read off the
left-most point in Fig.~\ref{fig_mbarSF}~\cite{mbar1}. The total
renormalization factor is thus known with a precision of 1.5\%.

\subsection{The strange quark mass in quenched QCD}

I shall now discuss the application of non-perturbative quark mass
renormalization to compute the mass of the strange quark in the
quenched approximation. Before discussing the details, it is useful to
recall that mass ratios of light quarks are predicted quite accurately
by Chiral Perturbation Theory (ChPT), for instance~\cite{Leutw96}
\be
   \frac{\Mstr}{\Mhat} = 24.4\pm1.4,\qquad
   \Mhat=\textstyle\frac{1}{2}(\Mup+\Md).
\label{eq_MChPT}
\ee 
In order to determine $\Mstr$, it is then sufficient to compute the
sum $(\Mstr+\Mhat)$ on the lattice and combine it with~\eq{eq_MChPT}.

\begin{figure}[tb]
\vspace{-4.5cm}
\hspace{2.cm}
\ewxy{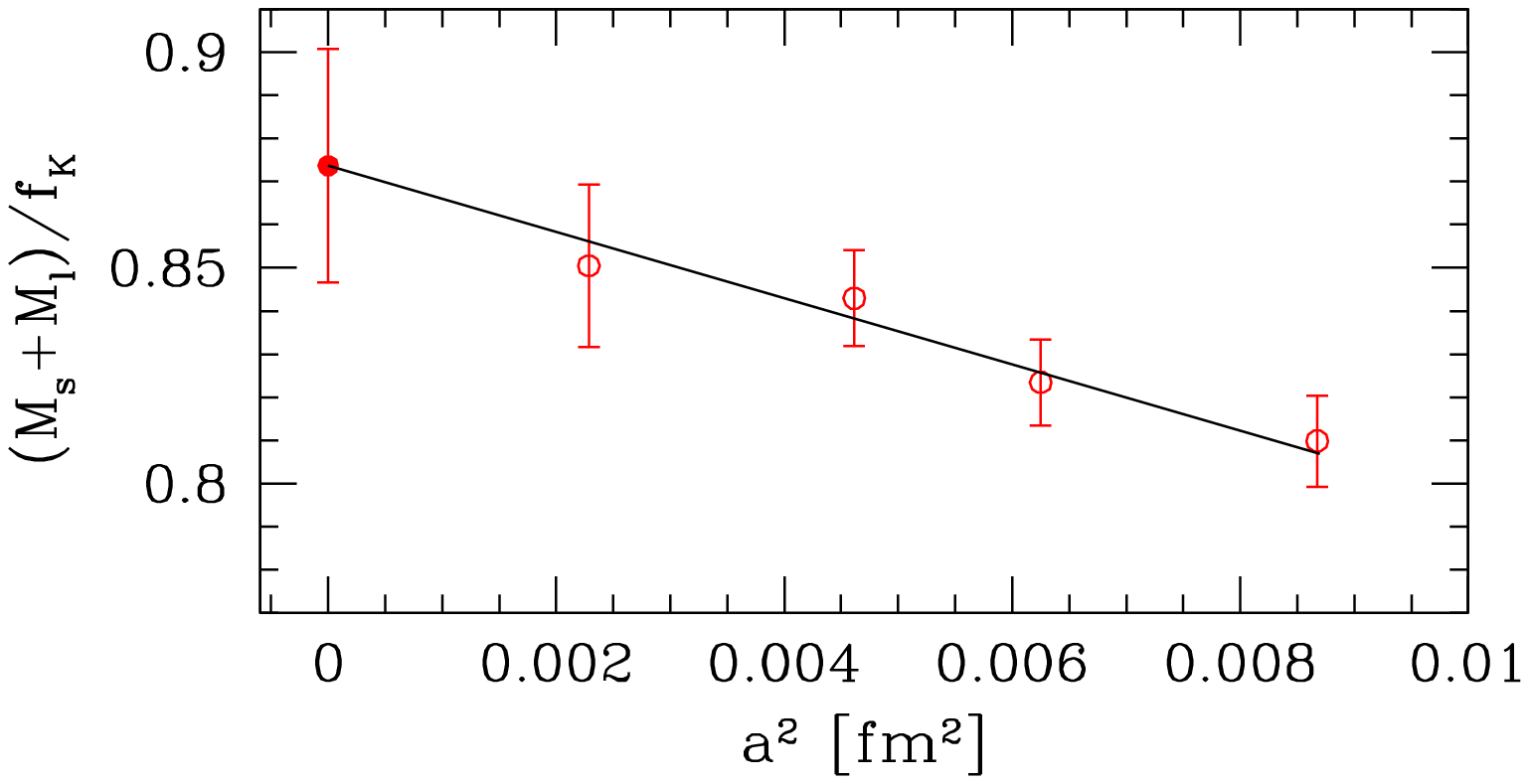}{120mm}
\vspace{-3.2cm}
\caption{Continuum extrapolation of $(\Mstr+\Mhat)/{\fK}$ from
  ref.~\protect\cite{mbar3}.}
\label{fig_MsMlcont}
\end{figure}

Here I shall concentrate on the calculation performed by the
ALPHA/UKQCD Collaboration~\cite{mbar3}. By combining the PCAC
relation, \eq{eq_PCAC}, with the total renormalization factor, one
obtains the sum of RGI quark masses $(\Mstr+\Mhat)$ in units of the
kaon decay constant\footnote{Here, $\ell$ denotes the Dirac spinor
  for the isospin-symmetric light quark.}
\be
   \frac{\Mstr+\Mhat}{\fK} = \left(\frac{M}{\mbar_{\rm SF}}\cdot
   \frac{1}{\zp}\right)\times
   \frac{\mK^2}{\langle0|\overline{\ell}\gamma_5{s}|{\rm
   K}\rangle_{\rm lat}} +O(a^2).
\label{eq_MfK}
\ee
The dependence on the lattice spacing can be eliminated by performing
a continuum extrapolation as shown in Fig.~\ref{fig_MsMlcont}.
Using the experimental value $\fK=160\pm2\,\mev$~\cite{PDG2000} one
obtains the result for the sum of RGI quark masses in the continuum
limit:
\be
   \Mstr+\Mhat = 140\pm5\,\mev.
\label{eq_MsMlres}
\ee
This result can now be converted into $\mbars^\MSbar(\bar{\mu})$ at
$\bar{\mu}=2\,\gev$. First, \eq{eq_MsMlres} is combined with the
prediction from ChPT, \eq{eq_MChPT}. Then, the relation between the
running mass in the $\MSbar$ scheme and the RGI quark mass is computed
in 4-loop perturbation theory as
\be
   \mbar^{\MSbar}(\bar{\mu})/M = 0.7208\;\;\hbox{\rm at}\;\;
  \bar{\mu}=2\,\gev.
\label{eq_mbarM_MSbar}
\ee
This procedure yields the final result in the quenched
approximation~\cite{mbar3}
\be
   \mbars^\MSbar(2\,\gev) = 97\pm4\,\mev.
\label{eq_ms_alpha}
\ee
The quoted uncertainty of $\pm4\,\mev$ contains all errors, except
those due to quenching. The high precision of this result is a direct
consequence of recent progress in lattice calculations, in particular
the implementation of non-perturbative renormalization and control
over lattice artefacts. As mentioned in section~\ref{sec_basic}, the
conversion into physical units is ambiguous in the quenched
approximation. For $\mbars^\MSbar(2\,\gev)$ the resulting uncertainty
was estimated in ref.~\cite{mbar3} to amount to $\sim10\%$.

\Table{ 
  Estimates for the quark masses $\mstr$ and $\mhat$ in the $\MSbar$
  scheme in quenched QCD. The choice of intermediate renormalization
  scheme~X is also shown, where applicable.
  \label{tab_quarkmass}
  } \br
Collaboration & $\mhat(2\,\gev)$ & $\mstr(2\,\gev)$ & X & $a\to0$ \\
\mr
CP-PACS \cite{CP-PACS_Nf2}          & 4.4(2)  & 110(4)   &    & yes \\
Becirevic \etal\,\cite{BGLM99}      & 4.8(5)  & 111(9)   & RI & no \\
G\"ockeler \etal\,\cite{QCDSF_quark}& 4.4(2)  & 105(4)   & SF & yes \\
Wingate \etal\,\cite{RBC_DWF}   &          & 130(11)(18) & RI & no \\
ALPHA/UKQCD \cite{mbar3}        &          & ~97(4)  & SF & yes \\
Blum \etal\,\cite{BSW_quark}    &          & ~96(26) &    & yes \\
JLQCD \cite{JLQCD_quark_stag}   & 4.23(29) & 106(7)  & RI & yes \\
Becirevic \etal\,\cite{Becir98} & 4.5(4)   & 111(12) & RI & no \\
Gim\'enez \etal\,\cite{Gimen98} & 5.7(1)(8)& 130(2)(18) & RI & no \\
\br \endTable

Table~\ref{tab_quarkmass} contains a compilation of recent results for
$\mstr$ and $\mhat$ obtained in the quenched approximation. Direct
comparisons of these results should be made with care, since
systematic errors have not been estimated in a uniform manner. Also,
the conversion into physical units has been performed using different
quantities. This manifests itself in the typical spread of the central
values, which is of the order of 10\%, consistent with the above
estimate. It is still quite remarkable that lattice estimates for
light quark masses have stabilized, which represents a big improvement
with respect to the situation before 1998.

\subsection{Sea quark effects in $\mstr$ and $\mhat$}

So far the most comprehensive study of quark masses using simulations
with dynamical quarks has been presented recently by
CP-PACS~\cite{CP-PACS_Nf2}. Earlier results can be found in
refs.~\cite{SESAM_quark97,SESAM_nf2_98}. When discussing the results,
it is important to keep in mind that non-perturbative renormalization
has not yet been implemented for $\Nf=2$ flavours. Instead the
renormalization of the lattice matrix element is usually performed
using so-called ``mean-field improved'' lattice perturbation
theory~\cite{lepenzie93}. In addition to computing the strange quark
mass, CP-PACS have also estimated the average light quark mass through
extrapolations to the chiral limit. Furthermore they have compared
different definitions of the quark mass, based either on the axial
vector Ward identity (AWI, i.e. the PCAC relation) or the vector Ward
identity (VWI).

\begin{figure}[tb]
\vspace{-3.3cm}
\hspace{3.cm}
\ewxy{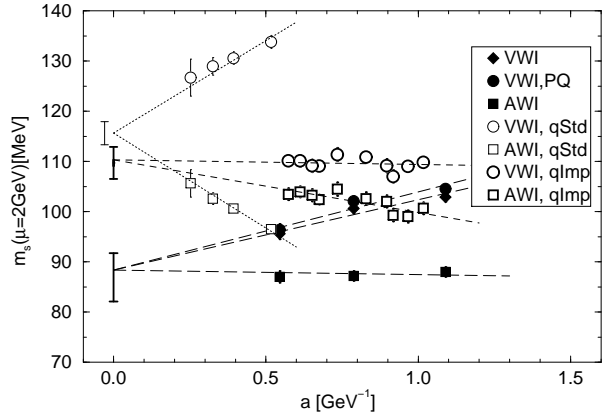}{100mm}
\vspace{-1.7cm}
\caption{Continuum extrapolations of the strange quark mass computed
  for $\Nf=2$ (solid symbols) and in the quenched approximation (open
  symbols). A different fermionic discretization used in the quenched
  case is distinguished by the bold open symbols.
  \label{fig_mstr_nf2}}
\end{figure}

The continuum extrapolations for the strange quark mass are shown in
Fig.~\ref{fig_mstr_nf2}. Results from the AWI and VWI definitions of
the quark mass have been extrapolated enforcing a common continuum
limit. By comparing the results for $\Nf=2$ to those obtained in the
quenched approximation, one finds that the inclusion of dynamical
quark effects decreases the estimate for the strange quark mass by
roughly 20\%.  The results for $\mstr$ by CP-PACS can be summarized as
follows:
\be
    \mstr^\MSbar(2\,\gev) = \left\{
    \begin{array}{rl}  88^{+4}_{-6}\,\mev,\quad & \Nf=2 \\
                      110^{+3}_{-4}\,\mev,\quad & \hbox{quenched\,.}
    \end{array}\right.
\ee
The light quark mass $\mhat$ shows a similar decrease when sea quark
effects are included. However, for both $\Nf=2$ and $\Nf=0$ (quenched)
one finds that $\mstr/\mhat\approx25$ in the continuum limit, which is
consistent with the prediction from ChPT, \eq{eq_MChPT}.

Figure~\ref{fig_mstr_nf2} also illustrates that lattice artefacts are
quite substantial, so that an estimation of quark masses at non-zero
lattice spacing would be misleading. This underlines once more the
importance of the continuum extrapolation.

\section{Concluding remarks \label{sec_concl}}

Lattice simulations of quenched QCD have reached a level of precision
of a few per cent, thanks to increases in computer power and
conceptual advances. The picture that has emerged is that the quenched
approximation works surprisingly well, as signified by the success in
predicting the spectrum of low-lying hadrons. Non-perturbative
renormalization and good control over lattice artefacts are the
crucial ingredients that lead to precise estimates of the mass of the
strange quark. Also, in order to control the region of very light
quarks, which cannot be simulated on present machines, it is useful to
combine lattice simulations with Chiral Perturbation Theory. In fact,
a combination of these techniques may be instrumental in resolving the
long-standing question of whether the up-quark is
massless~\cite{CoKapNel99}. A pilot study along these lines has
already appeared in the literature~\cite{mbar4}.

A major share of computational resources is devoted to studying the
effects of dynamical quarks. First results indicate that sea quarks
``are doing the right thing'', since the discrepancies in the light
hadron spectrum between simulation and experiment are decreased.
However, more effort is required to understand sea quark effects on a
quantitative level. This not only requires larger computers but also
the implementation of additional technology such as non-perturbative
renormalization for the dynamical case. Finally, there are algorithmic
challenges: the efficient simulation of sea quarks for which
$\mps/m_{\rm V}\ll0.5$, and also of odd~$\Nf$ is crucial for the
development of lattice QCD into a fully fledged phenomenological tool.

\ack I am grateful to the organizers, in particular Dr~Grazyna
Odyniec, for the invitation and the offered support. I wish to thank
all organizers for creating such a stimulating and pleasant
environment at the conference.


\section*{References}

\begin{thebibliography}{99}

\bibitem{Wilson74}
Wilson K\,G 1974 \PR {\bf D10} 2445

\bibitem{proc_lat99}
Campostrini M \etal (ed) 2000 {\it Proc. Int.
Symp. on Lattice Field Theory (Lattice\,99)} \NP~{\em B
(Proc. Suppl.)} {\bf 83-84}

\bibitem{kenway00}
Kenway R\,D 2000 plenary talk {\it Int. Conf. on High-Energy Physics
  (ICHEP\,2000) Osaka.} 

\bibitem{aoki99}
Aoki~S 1999 {\it Proc. Int. Symp. on
Lepton and Photon Interactions at High Energies (LP\,99)
Stanford, Preprint} hep-ph/9912288

\bibitem{eps99}
Wittig~H 1999, plenary talk {\it Int. Europhysics Conf. on
High-Energy Physics (EPS-HEP\,99)
Tampere}, {\em Preprint} hep-ph/9911400

\bibitem{sharpe_ichep98}
Sharpe~S 1998 {\it Proc. Int. Conf. on High-Energy Physics (ICHEP\,98)
Vancouver} vol~1 p~171 {\em Preprint} hep-lat/9811006

\bibitem{CP-PACS_quen}
CP-PACS Collaboration (Aoki~S \etal) 2000 \PRL {\bf 84} 238

\bibitem{GF11_quen}
Butler~F, Chen~H, Sexton~J, Vaccarino~A and Weingarten~D 1994 \NP 
{\bf B430} 179 

\bibitem{MILC_quen}
MILC Collaboration (Bernard~C \etal) 1998 \PRL {\bf 81} 3087

\bibitem{KS75}
Kogut~J and Susskind~L 1975 \PR {\bf D11} 395

\bibitem{UKQCD_quen}
UKQCD Collaboration (Bowler~K\,C \etal) 2000 \PR {\bf D62} 054506

\bibitem{SESAM_nf2_98}
SESAM Collaboration (Eicker~N \etal) 1999 \PR {\bf D59} 014509

\bibitem{UKQCD_c176}
UKQCD Collaboration (Allton~C\,R \etal) 1999 \PR {\bf D60} 034507

\bibitem{joyce_lat99}
UKQCD Collaboration (Garden~J \etal) 2000 \NP~{\em B (Proc. Suppl.)}
{\bf 83-84} 165

\bibitem{kaneko_lat99}
CP-PACS Collaboration (Ali~Khan~A \etal) 2000 \NP~{\em B (Proc. Suppl.)}
{\bf 83-84} 176

\bibitem{colum_00}
Chen~P \etal 2000
The finite temperature QCD phase transition with domain wall
fermions {\em Preprint} hep-lat/0006010.

\bibitem{PDG2000}
Particle Data Group (Groom~D\,E \etal) 2000 {\it Eur. Phys. J.} {\bf C3} 1

\bibitem{MPSTV94}
Martinelli~G, Pittori~C, Sachrajda~C\,T, Testa~M and Vladikas~A
1995 \NP {\bf 445} 81

\bibitem{LNWW92}
L\"uscher~M, Narayanan~R, Weisz~P and Wolff~U 1992 \NP {\bf B384} 168

\bibitem{sint_SF}
Sint~S 1994 \NP {\bf B421} 135; 1995 \NP {\bf B451} 416

\bibitem{martin_LesHouches}
L\"uscher~M 1998 Advanced Lattice QCD {\em Preprint} hep-lat/9802029

\bibitem{mbar1}
Capitani~S, L\"uscher~M, Sommer~R and Wittig~H
1999 \NP {\bf B544} 669

\bibitem{Leutw96}
Leutwyler~H 1996 \PL {\bf B378} 313

\bibitem{mbar3}
ALPHA \& UKQCD Collaborations (Garden~J \etal)
2000 \NP {\bf B571} 237

\bibitem{CP-PACS_Nf2}
CP-PACS Collaboration (Ali~Khan~A \etal) 2000 \PRL {\bf 85} 4674

\bibitem{BGLM99}
Becirevic~D, Gim\'enez~V, Lubicz~V and Martinelli~G
2000 \PR {\bf D61} 114507

\bibitem{QCDSF_quark}
G\"ockeler~M \etal 2000 \PR {\bf D62} 054504

\bibitem{RBC_DWF}
Wingate~M \etal 2000 \NP~{\em B (Proc. Suppl.)} {\bf 83-84} 221

\bibitem{BSW_quark}
Blum~T, Soni~A and Wingate~M 1999 \PR {\bf D60} 114507

\bibitem{JLQCD_quark_stag}
JLQCD Collaboration (Aoki~S \etal) 1999 \PRL {\bf 82} 4392

\bibitem{Becir98}
Becirevic~D \etal 1998 \PL {\bf B444} 401

\bibitem{Gimen98}
Gim\'enez~V, Giusti~L, Rapuano~F and Talevi~M 1999 \NP {\bf B540} 472

\bibitem{SESAM_quark97}
SESAM Collaboration (Eicker~N \etal) 1997 \PL {\bf B407} 290

\bibitem{lepenzie93}
Lepage~G\,P and Mackenzie~P\,B 1993 \PR {\bf D48} 2250

\bibitem{CoKapNel99}
Cohen~A\,G, Kaplan~D\,B and Nelson~A\,E 1999 {\em J. High Energy
  Phys.}  JHEP11(1999)027

\bibitem{mbar4}
ALPHA Collaboration (Heitger~J, Sommer~R and Wittig~H) 2000
\NP {\bf B588} 377

\end{thebibliography}
\end{document}